\documentclass[twocolumn,showpacs,preprintnumbers,amsmath,amssymb]{revtex4}
\usepackage{graphicx}
\usepackage{dcolumn}
\usepackage{bm}

\begin{document}

\title {Nonlinear spin-polarized transport through a ferromagnetic domain wall}
\author {G. Vignale}
\affiliation
{Department of Physics and Astronomy,  University of Missouri, Columbia,
Missouri
65211}
\author{M. E. Flatt\'e}
\affiliation
{Department of Physics and Astronomy,  University of Iowa, Iowa City, Iowa
52242}
\date{\today}
\begin{abstract}
A domain wall separating two oppositely magnetized regions in a ferromagnetic
semiconductor  exhibits, under appropriate conditions, strongly non linear I-V
characteristics similar to those of a $p$-$n$ diode.  We study these
characteristics as
functions of  wall width and temperature.   As the width increases or the
temperature decreases, direct tunneling between the majority spin bands
reduces the
effectiveness of the diode.  This has important implications for the zero-field
quenched resistance of magnetic semiconductors and for the design of a recently
proposed spin  transistor.
\end{abstract} \pacs{}
\maketitle

It has  recently been reported that some doped semiconductors, such as
Ga$_{1-x}$Mn$_x$As  \cite{OhnoRTD} and Ti$_{1-x}$Co$_x$O$_2$ \cite
{Matsumoto},  undergo ferromagnetic  transitions at temperatures as high as
$110$ K
and  $300$ K  respectively,  while others
($n$-doped Zn$_{1-x}$Mn$_x$Se  \cite{BeMnZnSe}) are almost completely spin
polarized by the application of a relatively modest magnetic field.   These
findings have
raised hopes for the realization of  semiconductor-based magnetoelectronic
devices
\cite {review}.

In a ferromagnetic semiconductor,  the up- and down- spin components of
just {\it
one} carrier type are quite analogous to majority and minority carriers in
ordinary doped
semiconductors.  Accordingly, a domain wall separating two ferromagnetic
regions
with opposite magnetizations is the analogue of a $p$-$n$ junction, while two
consecutive domain walls correspond to a $p$-$n$-$p$ transistor.  In a
recent paper
\cite{FV} we have exploited this analogy to show that  nonlinear
amplification of a
spin-polarized charge current is indeed possible in the ``$p$-$n$-$p$"
configuration, and
can be controlled by a magnetic field or  a voltage applied to the ``base"
region
between the two domain walls.   However,  the analysis of  Ref. \cite{FV}
was based
on the assumption that the probability of a carrier flipping its spin while
crossing the
domain wall is negligible.  This corresponds to assuming the resistivity of
the domain wall is large compared to that of the bulk material.

The resistance of a domain wall between ferromagnetic materials has been
examined several times from different perspectives since the pioneering
work of Cabrera and Falicov \cite{Falicov}. These authors found that the
resistance
was  very small, and later calculations \cite{Zhang,Simanek}
have supported that result for metallic magnets. A far different regime
is possible, however, when the spin polarization is or approaches $100 \%$.
For example, experimental and theoretical results\cite{Mathur} indicate
that domain
walls in La$_{0.7}$Ca$_{0.3}$MnO$_3$ may dominate  the resistance in thin
films.
Magnetic semiconductor systems, due to their very small bandwidths, are
also likely to
be $100 \%$ spin polarized, and thus their domain walls should be  highly
resistive in
the absence of spin-flip transport processes across  them.

A key question that has not been addressed so far is how the nonlinear
current-voltage
(I-V) characteristics of the domain wall  are affected by spin-flip
processes as the width of the domain wall increases. Note that the width
of a  domain wall can now be directly measured \cite{magneticSTM} and, in
principle,
geometrically controlled \cite{Bruno}.    Our analytical theory of
transport across the
domain  wall should therefore be useful in designing devices with optimal
values of
the  controllable parameters. Certainly such a theory would be crucial to
understanding
the zero-field  quenched resistance and the low-field magnetoresistance of
magnetic
semiconductors as well as to the realization of the ``unipolar spin
transistor" proposed
in \cite{FV}.

Here we present a quantitative study of the nonlinear I-V characteristics
of a magnetic
domain wall.  The main issue is the competition between minority spin
injection, which
is responsible for the nonlinear spin-diode behavior, and majority spin
transmission,
which tends to suppress it.  We shall show that the latter dominates when
either the
temperature is low, or the domain wall is   thick.   Assuming that the
motion of carriers
through the domain wall is ballistic, we derive  analytic expressions for
the charge and
spin currents as functions of applied voltage, width of the domain wall,
and temperature.  We further identify a new transport  regime for
intermediate wall
thicknesses, in which carriers are  ballistically transported across the
domain wall
(characterized by  nonlinear charge currents), but most spin polarization
is lost.

Our model is schematically depicted in Fig. 1(a).  The two
ferromagnetic regions $F1$ and $F2$ are connected by a domain wall region
of width
$d$, $-d/2<x<d/2$.  The exchange field $B(x)$ has the form
\begin{equation}
\vec B (x) = B_0 [\cos \theta(x) \hat x + \sin \theta (x) \hat y],
\end{equation}
where $\hat x$, $\hat y$  are  unit vectors in the direction of $x$ and
$y$, and the
angle $\theta (x)$ varies
linearly from $\theta = \pi/2$  in $F2$ to
 $\theta = -\pi/2$ in $F1$ \cite{footnote1}.

\begin{figure}
\label{fig1}
\includegraphics[width=7cm]{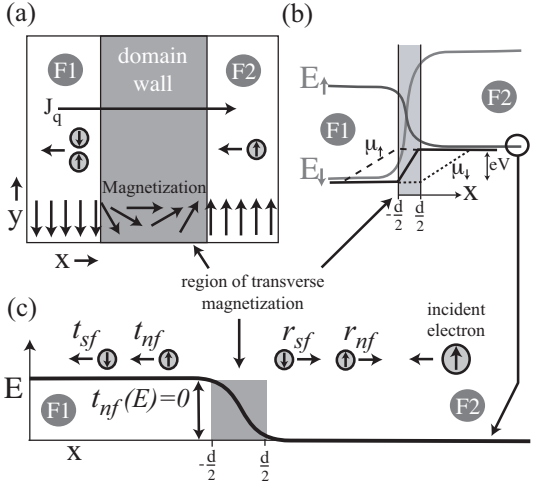}
\caption {(a) Schematic representation of a domain wall.  (b) Qualitative
behavior of
the quasichemical potentials and  the electrostatic potential (solid line).
Note that the
nonequilibrium voltage drop  occurs within the interfacial region, while the
nonequilibrium populations extend  up to a distance of order $L_s$  from
it. (c)
Reflection and transmission processes for an electron incident on the
domain wall. }
\vspace{-10pt}
\end{figure} 

We assume that  $d$, while possibly large in comparison  to a typical carrier
wavelength,  is  smaller than the mean free path and the spin diffusion length
$L_s$, which is  in turn smaller than the geometric size of the system. A
charge current
$J_q$ is injected from the left: our objective is to calculate the voltage
$V$ that
develops across the domain wall and the spin current $J_s$  due to the
flow.

Let $\mu_>$ and $\mu_<$ be the quasi-chemical potentials,
which control the nonequilibrium densities of majority and minority spin
carriers
respectively \cite{footnote2}.  Far from the wall we have  $\mu_> = \mu_<$
and the carrier densities have the equilibrium values $n^{(0)}_>$ and
$n^{(0)}_<$,
with $n^{(0)}_>  >> n^{(0)}_<$.    Density variations from equilibrium $\Delta
n_{>(<)} \equiv n_{>(<)}  - n^{(0)}_{>(<)} $ are related to the difference
of the
quasichemical potentials $\Delta \mu  \equiv \mu_< - \mu_>$ near the domain
wall.
Since, by charge neutrality,  $\Delta n_< \simeq - \Delta n_>$ we see
that the {\it relative} change in the minority spin  density is always much
larger than
the corresponding relative change in the majority spin density.   This
implies that
$\mu_>$ is essentially pinned to its bulk value, while $\mu_<$ varies
significantly in
a region of length $\sim L_s$ on either side of the domain wall.   We can
therefore set
$\mu_{>} \simeq 0$ throughout  $F1$ and $\mu_{>} \simeq  eV$ throughout  $F2$,
where $V$ is the electrostatic potential of $F1$ relative to $F2$ (see Fig.
1(b)) and the
carriers are assumed to be electrons.   The density variations are
 \begin{equation} \label{excessdensity}
 \Delta n_<  (x) = n^{(0)}_< \left [e^{\Delta \mu (x) /k_BT}-1 \right ],
\end{equation}
where $k_B$ is the Boltzmann constant and $T$ is the temperature.

The charge currents for majority and  minority spin orientations must
satisfy the
condition $J_> + J_< = J_q$ where the total charge current $J_q$ is
independent of
position.  In addition,  the minority carrier current $J_<$ is almost
entirely a diffusion
current, and is given by the classical relation $J_< (x) = e D  d n_< (x)
/dx$,  where
$D$ is the diffusion constant. Because the  spin density relaxes to equilibrium
exponentially on the scale of $L_s$ ( i.e., $\Delta n_{< }(x) =  \Delta
n_{<}(\pm d/2)
e^{-|x \mp d/2|/L_s}$ where the lower sign holds  in $F1$ and the upper
sign in $F2$),
the   minority carrier current at $x = \pm d/2$ can  be written as
$J_<(\pm d/2) = \mp
eD \Delta n_{<}(\pm d/2)/L_s$, or, with the help of Eq.~(\ref{excessdensity}),
\begin{equation} \label {minoritycurrents}
 J_<(\pm d/2) = \mp {eD n_<^{(0)} \over L_s} \left [e^{\Delta
\mu(\pm d/2)/k_BT}-1 \right ].
\end{equation}

It will be argued below that for nondegenerate carriers  the quasi-chemical
potential of minority spin  electrons on each side of the domain wall
adjusts to the
quasichemical potential of majority spin electrons on the opposite side, so
that
$\mu_<(-d/2) \simeq eV$, $\mu_<(d/2) \simeq 0$ (see Fig. 1(b)), and
\begin{equation}\label{deltamu}
\Delta \mu (\pm d/2) = \mp eV.
\end{equation}

Under the same assumption of nondegeneracy, it will also be shown that the
matching
condition for the spin current  $J_s (x) \equiv  J_\uparrow(x) -
J_\downarrow (x)$ is
\begin{equation} \label{matchingcondition}
{J_s(-d/2) \over
J_s(d/2)} =  {\bar t_{-} + \bar t_{+}e^{-eV/k_BT} \over  \bar t_{+} + \bar
t_{-}e^{-eV/k_BT}} \end{equation}  where $ \bar t_{\pm} =  \bar t_{nf} \pm \bar
t_{sf}$,   and $\bar t_{sf} $ and $\bar t_{nf}$  are population-averaged
transmission
coefficients, with  and without spin flip (see Fig. 1(c)), which will  be
defined more
precisely below.   Thus,  the spin current is conserved across a  sharp
domain wall
($\bar t_+ = \bar t_-$), but reverses its sign across a smooth one  ($\bar
t_+ = - \bar
t_-$).

Combining Eqs. (\ref{minoritycurrents}-\ref{matchingcondition}), and  using
current
conservation we arrive at our main results.  First
\begin{equation} \label{chargecurrent}
 { J _q\over J_0} = \sinh \left ( {e V \over k_B T} \right ) \left [1+
{\bar t_{sf }\over \bar t_{nf}} \tanh^2  \left ( {e V \over 2 k_B T} \right
) \right ],
\end {equation}
 where $J_0  \equiv 2 e D n^{(0)}_</L_s$.  For $\bar t_{sf} = 0$ this
reduces to the
equation \cite{Streetman} derived in (\cite {FV}), while for
$\bar t_{nf}=0$ we get  $V=0$ as expected for a ballistic conductor.   In
the linear
regime $eV/k_BT<<1$ this formula leads to  the well-known interfacial
resistance
of Fert and Valet \cite {Fert}.  Second, in the immediate vicinity of the
domain wall the
spin current is given by
\begin{equation} \label{spincurrent}
 { J_s \over J_0} = 2\sinh^2 \left ( {e V \over 2 k_B T} \right ) \left [1 \pm
{\bar t_{sf }\over \bar t_{nf}} \tanh  \left ( {e V \over 2 k_B T} \right )
\right ],
\end {equation}
where the upper sign holds in $F2$ and the lower sign in $F1$.  We see that
spin-flip processes cause the appearance of an odd-in-voltage component of the
spin-current,  whereas, for $t_{sf}=0$, the spin-current is  an even
function of $V$
\cite{FV}.   Shown in Fig. 2 is (a) the spin current in $F1$, (b) the
charge current,
and (c) the ratio of the two. The curves correspond to several different
values of  $\bar t_{nf}/ \bar t_{sf}$. The trends for the spin and charge
current
described above are evident  in Fig. 2; specifically the charge current is
always odd in
$V$ whereas  the spin current is even in the absence of spin-flip. When
spin-flip
dominates the spin current becomes odd as well. The spin current in $F2$ is
related to that in $F1$ according to the following relation: $J_s(F2;
V) = -J_s(F1; -V)$.  As $\bar t_{nf}/ \bar t_{sf} $ becomes
smaller, the  ``leakage current" between the two majority bands becomes
significant, and the odd in $V$ term in the spin
current begins to dominate.
Over the entire range shown of $\bar t_{nf}/ \bar
t_{sf} $ the relationship between $J_q$ and $V$ is highly
nonlinear indicating ballistic transport. Thus ballistic transport
itself is not a sufficient condition for maintaining spin polarization
in transport across a domain wall.

\begin{figure} \label{fig2}
\includegraphics[width=7cm]{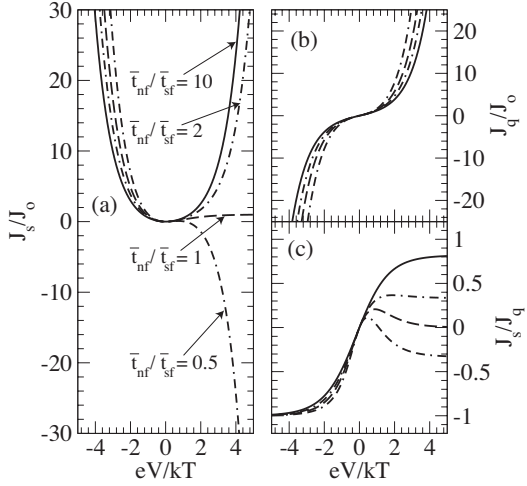}
\caption {(a)  Spin current in $F1$, (b) charge current, (c) ratio of
spin current to charge current vs.  voltage   for $\bar t_{nf}/ \bar
t_{sf}=10,\ 2,\ 1,\ 0.5.$}
\vspace{-10pt}
\end{figure}

Assuming ballistic transport in the wall region, we calculate the
transmission/reflection
coefficients from the exact  numerical solution of the Schr\"odinger equation
\begin{equation} \label{SE}
\left [ {- \hbar^2 \over 2m} {\partial^2 \over \partial x^2} - {\Delta
\over 2} \left (
\begin{array} {cc}  0 & e^{-i \theta (x)} \\ e^{+i \theta (x)}& 0 \end{array}
\right ) \right ]
\left ( \begin {array} {cc}\psi_{\uparrow} \\ \psi_\downarrow \end{array}
\right )
=  E \left ( \begin {array}{cc} \psi_{\uparrow} \\ \psi_\downarrow
\end{array} \right
),
\end{equation}
where $\Delta = g \mu_B B$ is the exchange spin-splitting.  The technique of
solution is the same as used in Ref. \cite{Zhang}.
Sample results are shown in Fig.
3(a)-(c) for three different values of the dimensionless  parameter $\xi =
\hbar \pi/2 d
\sqrt {2 m \Delta} = 10$, $1$, and $0.1$,  corresponding to sharp,
intermediate,  and
smooth domain walls respectively.  smooth domain walls respectively.
Recent experiments \cite{magneticSTM} suggest
the width of domain walls in artificial nanostructures can be as small as
$1$~nm, giving $\xi \sim 1$ for an effective mass $m$ equal to the electron
mass and a spin splitting   $\Delta = 100 meV$.
Domain walls thinner than $20$~nm have already been inferred in thin
GaMnAs layers\cite{Schiffer}.

\begin{figure} \label{fig3}
\includegraphics[width=7cm]{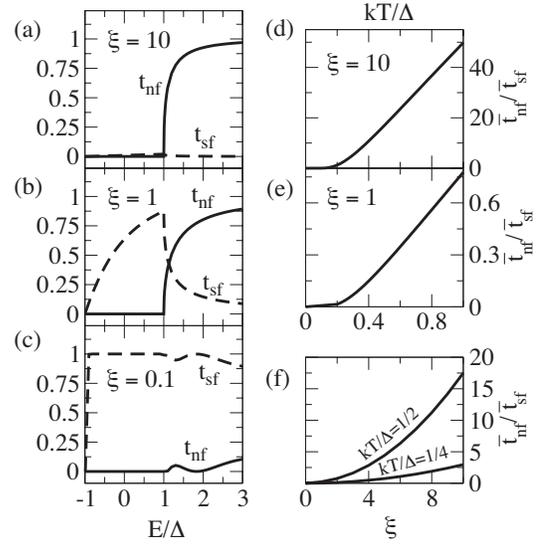}
  \caption{(a)-(c) Energy dependence of transmission coefficients for
 $\xi =10$, $1$, and $0.1$  respectively . (d)-(e) Ratio of the
population averaged non-spin-flip to  spin-flip transmission
coefficients ($\bar t_{nf}/ \bar t_{sf}$) vs.
temperature for $\xi = 10$ and $1$ respectively. (f) $\bar t_{nf}/ \bar
t_{sf}$ versus $\xi$ for $k_BT/\Delta=0.25$ and $0.5$.}
\vspace{-10pt}
\end{figure}

Fig. 3(d)-(f) shows the behavior of the key ratio $\bar
t_{nf}/ \bar t_{sf}$ as a function of temperature and thickness.  As expected
$\bar t_{nf}$ vanishes at low temperature, because, in this limit, there
are no incident states above the exchange barrier to provide  minority
spin-injection.  The spin diode is a thermally-activated device (as a
$p$-$n$ diode is), thus higher temperature is favorable to
its performance.   Fig. 3(d,e) supports this view by
showing that minority spin injection only dominates above a certain
temperature (depending on domain wall thickness).   However the
condition $k_BT \lesssim \Delta$ must be respected if the system is to be
nearly
$100 \%$ spin-polarized. The conclusion is that there is a range
$T_{min}<T<T_{max}$ in which unipolar spin diodes and transistors are
expected to
be  operational.

 We now come to the
justification of the matching condition (\ref{matchingcondition}) and the
calculation of the quasi-chemical potential offset.   We begin with the
former.   In the
spirit of the Landauer-B\"uttiker formalism we treat the ferromagnetic
regions $F1$ and
$F2$ as two reservoirs of spin polarized electrons at  chemical potentials
$\mu_1 = 0$ and $\mu_2 = eV$  which inject
up- and down-spin electrons, respectively, in the domain wall region.  The
small
density of minority spin carriers is neglected in the following argument.  The
components of the current due to electrons with energies in the range
$(E, E+dE)$ on the two sides of the domain wall are given (in units of $e/h$) by
 \begin{eqnarray} \label{LBcurrents}
j_{1 >}(E) &=& -(1 - r_{nf}(E)) f
_{1>}(E) + t_{sf}(E)f_{2>} (E)
\nonumber \\
j_{1 <} (E) &=&  ~~r_{sf}(E)f_{1>}(E) + t_{nf}(E)f_{2>}(E)
\nonumber  \\
j_{2 >} (E) &=&  ~~(1 - r_{nf}(E)) f_{2>}(E) - t_{sf}(E)f_{1>}(E)
\nonumber \\
j_{2 <} (E) &=&  ~- r_{sf}(E)f_{2>}(E) - t_{nf}(E)f_{1>}(E),
 \end{eqnarray}
where $r_{nf}$ and $r_{sf}$ are the non spin-flip and spin-flip
reflection probabilities, related to $t_{nf}$ and $t_{sf}$  by the
unitarity condition
$r_{nf}+r_{sf}  +t_{nf}+t_{sf} = 1$, and $f_{1 >}$ ,$f_{2 >}$ are
shorthands for the
equilibrium distributions of majority spin carriers in $F1$ and $F2$
respectively.
Note that, for nondegenerate carriers $f_{1>} =  f_{2>} e^{-eV/k_BT}$.
We find that the spin-flip reflection coefficient $r_{sf}$ is extremely
small at all
energies and thicknesses, and can therefore be safely neglected.  With this
approximation, combined with the unitarity condition, it is easy to show
that the
energy-resolved  currents are given by  $j_{s1(2)} (E) = ( t_{-(+)} (E) +
t_{+(-)} (E)
e^{-eV/k_BT}) f_{2>}(E)$. Noting that $f_{2>}(E)  \propto
~e^{-E/k_BT}$ and integrating over  energy we see that the total current
$J_{s1} =
\int_0^\infty j_{s1}(E) e^{-e/k_BT}$ is equal to $A ( \bar t_- + \bar t_+
e^{-eV/k_BT})$ where the average transmission coefficients are defined as
\begin{equation}
\label{averagetransmission}
\bar   t_{nf (sf)} = {\int_0^\infty  t_{nf (sf)}(E) e^{-E/k_BT} dE \over
\int_0^\infty e^{-E/k_BT} dE},
\end{equation}
and $A$ is a constant. Similarly  $J_{s2} =  A( \bar t_+ + \bar t_-
e^{-eV/k_BT})$.  The ratio $J_{s1}/J_{s2}$ is thus given by
Eq.~(\ref{matchingcondition}).

To justify the quasi-chemical potential offset condition,
Eq.~(\ref{deltamu}) we notice that the quasi-chemical potential
$\mu_{<,1}$ of minority spin electrons near the left  hand side of the
domain wall is an
average of the quasi-chemical potentials of right (+) and left (-) moving
electrons :
$e^{ \mu_{<,1}/k_BT} = [e^{ \mu_{<,1}^+/k_BT}+e^{ \mu_{<,1}^-/k_BT}]/2$.  (A
similar relation holds for the quasi-chemical potential  $\mu_{<,2}$ of
minority spin
electrons near the right hand side of the domain wall). The quasi-chemical
potentials for
right and left movers on either side are determined by the conditions of
continuity
\begin{eqnarray}
\label{chemicalpotentials}  e^{-(E - \mu_{>,2}^+)/k_BT } &=& q e^{-(E -
\mu_{>,1}^+)/k_BT }  + p e^{-(E - \mu_{<,1}^+)/k_BT } \nonumber \\
e^{-(E - \mu_{<,2}^+)/k_BT } &=& q e^{-(E - \mu_{<,1}^+)/k_BT }  + p
e^{-(E - \mu_{>,1}^+)/k_BT }
\nonumber \\
e^{-(E - \mu_{<,1}^-)/k_BT } &=& q e^{-(E - \mu_{<,2}^-)/k_BT }  + p
e^{-(E - \mu_{>,2}^-)/k_BT }
\nonumber \\
e^{-(E - \mu_{>,1}^-)/k_BT } &=& q e^{-(E - \mu_{>,2}^-)/k_BT }  + p
e^{-(E - \mu_{<,2}^-)/k_BT }, \nonumber \\
\end{eqnarray}
where $q = t_{sf}/(t_{sf}+t_{nf})$ and $p = t_{nf}/(t_{sf}+t_{nf})$ are the
relative probabilities of transmission with and without spin flip respectively.
The first of these equations, for example,  says that the density of
right-moving
up-spin electrons of energy $E$ on the right hand side of the domain wall
is equal
to  the density of   right-moving down-spin electrons of the same energy
which enter
from the left and flip their spin, {\it plus}  the density of
right-moving up-spin
electrons which enter from the left and do not flip their spin.   Because the
quasi-chemical potentials of the majority spin carriers are essentially
pinned to their
bulk values, we can set $\mu_{>,2}^+ = \mu_{>,2}^- = \mu_{>,2}=eV$ and
$\mu_{>,1}^+ = \mu_{>,1}^- = \mu_{>,1}=0$.  Integrating
Eqs.~(\ref{chemicalpotentials}) over energy, and making use of $p+q = 1$,
we easily
get $\mu_{<,1}=eV$ and  $\mu_{<,2}=0$, as indicated in Fig. 1(b).

In summary, we have shown that both the  thickness and the temperature
have a profound influence on the nonlinear transport properties of a
ferromagnetic
domain wall.  We have derived  analytical formulas, Eqs.
(\ref{chargecurrent}) and (\ref{spincurrent}), for the charge and spin
currents of this
``magnetic junction" under physical  assumptions similar to the ones from
which the
Shockley equations of a  classical $p$-$n$ junction are derived. These formulae
indicate a new  transport regime, where charge transport is ballistic,  but
spin
polarization is lost.  Equations (\ref{chargecurrent}) and
(\ref{spincurrent}), together
with  microscopic calculation of the population-averaged transmission
coefficients,
can be used to assess the effectiveness of unipolar spin-diode devices in
realistic
circumstances.

We gratefully acknowledge support from   NSF grants No. DMR-0074959
and from DARPA/ARO DAAD19-01-1-0490.
%***********************************************


\begin{thebibliography}{}
\bibitem{footnote2}  The exact electronic states are, strictly
speaking,  neither ``up" nor ``down"  with respect to the local
magnetization, but
superpositions of the two orientations. However,  the relative phase of the
up and down
components of the wave function   has a rapid spatial variation ---
typically on the
scale of the electronic wavelength  $\lambda$.  Therefore transverse spin
coherence
disappears on a scale larger than $\lambda$, and the carriers' spin can be
safely
assumed to be either parallel or antiparallel to the local magnetization.
\end{thebibliography}
\end{document}